\documentclass[graybox]{svmult}

\usepackage{mathptmx}       
\usepackage{helvet}         
\usepackage{courier}        
\usepackage{type1cm}        
\usepackage{epsfig,amsmath,amssymb,graphics} 
\usepackage{makeidx}         
\usepackage{graphicx}        
\usepackage{multicol}        
\usepackage[bottom]{footmisc}

\makeindex             

\begin{document}

\title*{Universality in Systems with Power-Law Memory  
and Fractional Dynamics}
\author{M. Edelman}
\institute{M. Edelman \at Department of Physics, Stern College at 
Yeshiva University\\ 
245 Lexington Ave, New York, NY 10016, USA; \\
Courant Institute of Mathematical Sciences, New York University \\
251 Mercer St., New York, NY 10012, USA; \\ 
Department of Mathematics, BCC, CUNY, \\
2155 University Avenue, Bronx, New York 10453, USA,\\
 \email{edelman@cims.nyu.edu}}

%
%
\maketitle

\abstract*{There are a few different ways to extend regular nonlinear dynamical
systems by introducing power-law memory or considering fractional
differential/difference equations instead of integer ones.
This extension allows the introduction of families of nonlinear dynamical
systems converging to regular systems in the case of an integer power-law
memory or an integer order of derivatives/differences. The examples
considered in this review include the logistic family of maps 
(converging in the case of the first order difference to 
the regular logistic map), the universal family of maps, and the standard 
family of maps (the latter two converging, in the case
of the second difference, to the regular universal and standard maps).   
Correspondingly, the phenomenon of transition to chaos through a 
period doubling cascade
of bifurcations in regular nonlinear systems, known as ``universality'',
can be extended to fractional maps, which are maps with
power-/asymptotically power-law memory. The new features of universality,
including cascades of bifurcations on single trajectories, which appear in
fractional (with memory) nonlinear dynamical systems are the main subject
of this review. 
}


\section{Introduction}
\label{sec:1}
Sir Robert M. May's paper ``Simple mathematical models with very complicated
dynamics'' published 41 years ago in ``Nature'' \cite{May} is one of the
most cited papers - 3057 citations are registered by the Web of Science 
at the moment I am writing this sentence. In this review the author, using the
logistic map as an example, described the universal behavior typical
for all nonlinear systems: transition to chaos through the period-doubling
cascade of bifurcations. The main applications considered by the author
are the biological (even the variable used in the text was treated as ``the
 population''), economic, and social sciences. The major
steps in the development of the notion of universality in non-linear
dynamics are gathered in the reprinted selection of papers compiled
by Predrag Cvitanovic \cite{Cvi}, and applications of 
universality encompass all areas of science.

The logistic map is a very simple discrete non-linear model of dynamical
evolution. More realistic models of biological, economic, and social
systems are more complicated. One of the features, not
reflected in this equation but which is present in all the abovementioned systems, 
is memory. Evolution of any social or biological system depends not
only on the current state of a system but on the whole history of its
development. In majority of cases this memory obeys the power law.
There are many reviews on power-law distributions and memory in various 
social systems (see, e.g., \cite{Mach}). In papers 
\cite{Adaptation3,Adaptation4,Adaptation2,Adaptation5,Adaptation1,Adaptation6} 
the power-law adaptation has been used to describe the dynamics of biological 
systems. The impotence and origin of the memory in biological systems can be 
related to the presence of memory at the level of individual cells:  
it has been shown recently that processing of external stimuli by individual neurons 
can be described by fractional differentiation \cite{Neuron3,Neuron4,Neuron5}. 
The orders of fractional derivatives $\alpha$ derived for different types of 
neurons fall within the interval [0,1], which implies power-law memory 
$\sim t^{\beta}$ with power $\beta = \alpha - 1$, $\beta \in [-1,0]$. 
For neocortical pyramidal neurons the order of the fractional derivative 
is quite small: $\alpha \approx 0.15$. 
At the level of a human individual as a whole the power law appears in the 
study of human memory: forgetting - the accuracy on memory tasks decays as a power law with $\beta \in [-1,0]$ 
\cite{Kahana,Rubin,Wixted1,Wixted2,Adaptation1}; 
learning - the reduction in reaction times that comes with
practice is a power function of the number of training trials \cite{Anderson}.
Power-law memory appears in the study of the human organ tissues
due to their viscoelastic properties (see, e.g., references in \cite{Chaos2015}). 
This leads to their description by fractional differential equations with
time fractional derivatives which implies the power-law memory. In most of the
biological systems with the power-law behavior the power $\beta$ is between -1 and 1 ($0 < \alpha <2$).
 
It is much easier to investigate general properties of discrete systems with power-law memory than properties of integro-differential equations with power-law kernel. In section \ref{sec:2} we review different ways to introduce or derive maps with power-law memory and their relation to fractional differential/difference equations. 
Periodic sinks and their stability (the stability of fixed points and asymptotic period two ($T=2$) sinks) in fractional systems are discussed in section \ref{sec:3}. In section \ref{sec:4} we consider various forms (non-linearity parameter, two-dimensional, and memory parameter) of bifurcation diagrams and transition to chaos in discrete fractional systems. In the conclusion we discuss perspectives and application of the research on the universality in fractional systems.

\section{Maps with power-law memory and fractional maps}
\label{sec:2}

In this section we consider various ways to introduce maps with power-law memory and fractional maps following \cite{ME2,ME3,ME4,ME6,ME7,ME8,ME9,Chaos2015,ME1,ME5, T2009a,T2009b,T2,T1,FallC,Fall}.

In the following we will use two definitions of fractional
derivatives. They are based on the fractional integral introduced by Liouville, which is a generalization of the Cauchy formula for the n-fold integral
{\setlength\arraycolsep{0.5pt}
\begin{equation}
 _aI^{p}_t x(t) 
=\frac{1}{\Gamma(p)} \int^{t}_a 
\frac{x(\tau) d \tau}{(t-\tau)^{1-p}}~,
\label{RLI}
\end{equation}
}
where $p$ is a real number, $\Gamma()$ is the gamma function and we will assume $a=0$.

The first one is the left-sided Riemann-Liouville fractional
derivative $_0D^{\alpha}_tx(t)$ 
defined for
$t>0$ \cite{KST,Podlubny,SKM} as 
{\setlength\arraycolsep{0.5pt}
\begin{equation}
_0D^{\alpha}_t x(t)=D^n_t \ _0I^{n-\alpha}_t x(t) 
=\frac{1}{\Gamma(n-\alpha)} \frac{d^n}{dt^n} \int^{t}_0 
\frac{x(\tau) d \tau}{(t-\tau)^{\alpha-n+1}}~,
\label{RL}
\end{equation}
}
where $n-1 \le \alpha < n$, $n \in \mathbb{Z}$,  
$D^n_t=d^n/dt^n$.        

The second one is the left-sided Caputo derivative, in which the order of integration and differentiation in Eq. (\ref{RL}) is switched
\cite{KST}  
{\setlength\arraycolsep{0.5pt}
\begin{equation}
_0^CD^{\alpha}_t x(t)=_0I^{n-\alpha}_t \ D^n_t x(t)
=\frac{1}{\Gamma(n-\alpha)}  \int^{t}_0 
\frac{ D^n_{\tau}x(\tau) d \tau}{(t-\tau)^{\alpha-n+1}}  \quad (n-1 <\alpha \le n).
\label{Cap}
\end{equation}
}

\subsection{Direct introduction of maps with power-law memory}
\label{Direct}

The direct way to introduce maps with power-law memory is to define them as convolutions according to the formula (see \cite{Chaos2015,StanislavskyMaps})
\begin{equation}
x_{n}=\sum^{n-1}_{k=0}(n-k)^{\alpha-1} G_K(x_k,h),
\label{LTMPL}
\end{equation}
where $K$ is a parameter and $h$ is a constant time step between 
time instants $t_n$ and $t_{n+1}$. For a physical interpretation of this formula we consider a system which state is defined by the variable $x$ and evolution by the function $G_K(x)$. The value of the state variable at the time $t_n$ is a weighted total of the functions $G_K(x_k)$ from the values of this variable at past time instants $t_k$, $0<k<n$, $t_k=kh$. The weights are the times between time instants $t_n$ and $t_k$ to the fractional power $\alpha-1$.

The more general form of this map considered in \cite{Chaos2015,StanislavskyMaps} (see, e.g., Eq.~(73) from \cite{Chaos2015}) is
\begin{equation}
x_{n}=\sum^{\lceil \alpha \rceil - 1}_{k=1}\frac{c_k}{\Gamma(\alpha-k+1)}
(nh)^{\alpha-k}  +\sum^{n-1}_{k=0}(n-k)^{\alpha-1} G_K(x_k,h),
\label{FrLTMPLNN}
\end{equation}
where $\alpha \in \mathbb{R}$. If we assume 
\begin{equation}
G_K(x,h)=\frac{1}{\Gamma (\alpha)}h^{\alpha}G_K(x), 
\label{FORxSmooth}
\end{equation}
where $G_K(x)$ is continuous, and \begin{equation}
x=x(t), \ \  x_k=x(t_k), \ \ t_k=a+kh, \ \ nh=t-a 
\label{GLDef}
\end{equation}
for $0 \le k \le n$, then this equation can be written as
\begin{equation}
x_{n}=\sum^{\lceil \alpha \rceil - 1}_{k=1}\frac{c_k}{\Gamma(\alpha-k+1)}
(nh)^{\alpha-k}  +\frac{h^{\alpha}}{\Gamma({\alpha})}\sum^{n-1}_{k=0}(n-k)^{\alpha-1} G_K(x_k).
\label{EqDir}
\end{equation}
Eq.~ (\ref{EqDir}) in the limit $h \rightarrow 0+$ 
will yield the Volterra integral equation of the second kind 
\begin{equation}
x(t)= \sum^{\lceil \alpha \rceil-1}_{k=1}\frac{c_k}{\Gamma(\alpha-k+1)}(t-a)^{\alpha-k}+\frac{1}{\Gamma (\alpha)}
\int^{t}_{a}\frac{G_K(\tau,x(\tau))d\tau}{(t-\tau)^{1-\alpha}}, \  \ \ (t>a). 
\label{VoltRealNN}
\end{equation}
This equation is equivalent to the fractional differential equation with the Riemann-Liouville or Gr$\ddot{u}$nvald-Letnikov fractional derivative \cite{Chaos2015,KBT1,KBT2}
\begin{equation}
_a^{RL/GL}D^{\alpha}_tx(t)=G_k(t,x(t)), \  \ 0 <\alpha
\label{KMRL}
\end{equation}
with the initial conditions 
\begin{equation}
(_a^{RL/GL}D^{\alpha-k}_tx) (a+)=c_k, \  \  \ k=1,2,...,\lceil \alpha \rceil.
\label{IC3}
\end{equation}
For $\alpha \not\in \mathbb{N}$ we assume $c_{\lceil \alpha \rceil}=0$, which corresponds to a finite value of $x(a)$.

\subsection{Universal map with power-law memory from 
fractional differential equations of systems with 
periodic delta-function kicks}
\label{SecUFM}

The universal map and its particular form, the standard map, play an
important role in the study of regular dynamical systems. Their fractional
generalizations can be obtained in a way similar to the way in which the
regular universal map is derived from the differential equation of a
periodically (with the period $h$) kicked system in regular dynamics (see, e.g., \cite{ZasBook}). The
two-dimensional fractional universal map obtained from the differential
equation of the order $1 < \alpha \le 2$ was introduced in \cite{T1},
extended to any real $\alpha >1$ in \cite{T2009a,T2009b,T2}, and then to
any $\alpha \ge 0$ in \cite{ME3,ME4,ME7}.

To derive the equations of the fractional universal map, which we'll call
the universal $\alpha$-family of maps ($\alpha$-FM) for $\alpha \ge 0$, 
we start with the differential equation 
\begin{equation}
\frac{d^{\alpha}x}{dt^{\alpha}}+G_K(x(t- \Delta h)) \sum^{\infty}_{k=-\infty} \delta \Bigl(\frac{t}{h}-(k+\varepsilon)
\Bigr)=0,   
\label{UM1D2Ddif}
\end{equation}
where $\varepsilon > \Delta > 0$,  $\alpha \in \mathbb{R}$, $\alpha>0$,
and consider 
it as $\varepsilon  \rightarrow 0$. The initial conditions
should correspond to the type of fractional derivative
used in Eq.~(\ref{UM1D2Ddif}).
The case $\alpha =2$, $\Delta = 0$, and $G_K(x)=KG(x)$  
corresponds to the equation whose integration yields the regular
universal map. 

Integration of Eq.~(\ref{UM1D2Ddif}) with the Riemann-Liouville fractional
derivative $_0D^{\alpha}_tx(t)$ and the initial conditions  
\begin{equation}
(_0D^{\alpha-k}_tx)(0+)=c_k, 
\label{ic}
\end{equation}
where $k=1,...,N$ and $N=\lceil \alpha \rceil$, yields the
Riemann-Liouville universal $\alpha$-FM 
\begin{equation}
x_{n+1}=  \sum^{N}_{k=1}\frac{c_k}{\Gamma(\alpha-k+1)}h^{\alpha -k}(n+1)^{\alpha -k}-\frac{h^{\alpha}}{\Gamma(\alpha)}\sum^{n}_{k=0} G_K(x_k) (n-k+1)^{\alpha-1}. 
\label{FrRLMapx} 
\end{equation}
As in the Sec.~\ref{Direct}, for $\alpha \not\in \mathbb{N}$  
boundedness of $x(t)$ at $t=0$ requires $c_N=0$ and
$x(0)=0$ (see \cite{KST,Podlubny,SKM}). 
Obtained in Sec.~\ref{Direct} Eq.~(\ref{EqDir}) is identical to Eq.~(\ref{FrRLMapx}).

Integration of Eq.~(\ref{UM1D2Ddif}) with the Caputo fractional derivative
$_0^CD^{\alpha}_t x(t)$ and the initial conditions
$(D^{k}_tx)(0+)=b_k$,  
$k=0,...,N-1$, yields the Caputo  universal $\alpha$-FM 
\begin{eqnarray}
x_{n+1}= \sum^{N-1}_{k=0}\frac{b_k}{k!}h^k(n+1)^{k} 
-\frac{h^{\alpha}}{\Gamma(\alpha)}\sum^{n}_{k=0} G_K(x_k) (n-k+1)^{\alpha-1}.
\label{FrCMapx}
\end{eqnarray}


 Later in this paper we'll refer to the maps 
Eqs.~(\ref{EqDir})~and~(\ref{FrRLMapx}), the RL universal $\alpha$-FM, as
the Riemann-Liouville universal map with power-law memory or 
the Riemann-Liouville universal fractional map; 
we'll call the Caputo universal  $\alpha$-FM, Eq.~(\ref{FrCMapx}),     
the Caputo universal map with power-law memory or 
the Caputo universal fractional map.

In the case of integer $\alpha$ the universal map converges to
\begin{equation}
x_n=0  \  \ {\rm for} \  \ \alpha = 0 \  \ {\rm and} \  \ x_{n+1}=x_n-hG_K(x_n)  
\  \ {\rm for} \  \  \alpha = 1, 
\label{IntFr01} 
\end{equation}
and for $\alpha=N = 2$ with $p_{n+1}=(x_{n+1}-x_n)/h $
\begin{equation}
\begin{array}{c}
\left\{
\begin{array}{lll}
p_{n+1}=p_n-h G_K(x_{n}), \ \ n \ge 0,
\\ 
x_{n+1} = x_{n}+hp_{n+1}, \ \ n \ge 0.
\end{array}
\right.
\end{array} 
\label{IntFr2}
\end{equation} 
N-dimensional, with  $N \ge 2$, universal maps are investigated in \cite{ME4},
where it is shown that they are volume preserving.

\subsection{Universal fractional difference map}
\label{SecUFDM}

The fractional 
sum ($\alpha>0$)/difference ($\alpha<0$) operator introduced by Miller and Ross in \cite{MR} 
\begin{equation}
_a\Delta^{-\alpha}_{t}f(t)=\frac{1}{\Gamma(\alpha)} \sum^{t-\alpha}_{s=a}(t-s-1)^{(\alpha-1)} f(s)
\label{MRDef}
\end{equation}
can be considered as a fractional generalization of the $n$-fold summation formula \cite{ME9,GZ}
\begin{equation} 
_a\Delta^{-n}_{t}f(t)=\frac{1}{(n-1)!} \sum^{t-n}_{s=a}(t-s-1)^{(n-1)}
f(s)
=\sum^{t-n}_{s^0=a} \sum^{s^0}_{s^1=a}...
\sum^{s^{n-2}}_{s^{n-1}=a}f(s^{n-1}),
\label{MRInt}
\end{equation}
where $n \in \mathbb{N}$ and $s^{i}$, $i=0,1,...n-1$, are the summation variables. In Eq.~(\ref{MRDef}) $f$ is defined on  $\mathbb{N}_a$ and $_a\Delta^{-\alpha}_t$ on  
$\mathbb{N}_{a+\alpha}$, where   $\mathbb{N}_t=\{t,t+1, t+2, ...\}$.
The falling factorial $t^{(\alpha)}$ is defined as
\begin{equation}
t^{(\alpha)} =\frac{\Gamma(t+1)}{\Gamma(t+1-\alpha)}, \ \ t\ne -1, -2, -3,
...
\label{FrFac}
\end{equation}
and is asymptotically a power function:
\begin{equation}
\lim_{t \rightarrow
  \infty}\frac{\Gamma(t+1)}{\Gamma(t+1-\alpha)t^{\alpha}}=1,  
\ \ \ \alpha \in  \mathbb{R}.
\label{GammaLimit}
\end{equation}

For $\alpha >0$ and   $m-1<\alpha < m$ the fractional (left) Riemann-Liouville difference operator is defined (see \cite{Atici1,Atici2}) as
\begin{equation} 
_a\Delta^{\alpha}_t x(t) =  \Delta^{m} _a\Delta^{-(m-\alpha)}_{t}x(t)
=\frac{1}{\Gamma(m-\alpha)} \Delta^m \sum^{t-(m-\alpha)}_{s=a}(t-s-1)^{(m-\alpha-1)} 
x(s)
\label{FDRL}
\end{equation}
and the fractional (left) Caputo-like difference operator (see \cite{Anastas}) as
\begin{equation} 
_a^C\Delta^{\alpha}_t x(t) =  _a\Delta^{-(m-\alpha)}_{t}\Delta^{m} x(t)
=\frac{1}{\Gamma(m-\alpha)} \sum^{t-(m-\alpha)}_{s=a}(t-s-1)^{(m-\alpha-1)} 
\Delta^m x(s),
\label{FDC}
\end{equation}
where $\Delta^{m}$ is the $m$-th power of the forward difference operator
defined as $\Delta x(t)=x(t+1)-x(t)$. 
Due to the fact that $_a\Delta^{\lambda}_t$ in the limit 
$\lambda \rightarrow 0$ approaches the identity operator (see \cite{ME9,MR}), the 
definition Eq.~(\ref{FDC}) can be extended to all real $\alpha \ge 0$
with $_a^C\Delta^{m}_t x(t) = \Delta^m x(t)$ for $m \in \mathbb{N}_0$.

Fractional h-difference operators, which are generalizations of the fractional difference operators, were introduced and investigated in \cite{hdif1,hdif2,hdif3,hdif4,hdif4n,hdif5,hdif6}. 
The h-sum operator is defined as
\begin{equation}
(_a\Delta^{-\alpha}_{h}f)(t)=\frac{h}{\Gamma(\alpha)} \sum^{\frac{t}{h}-\alpha}_{s=\frac{a}{h}}(t-(s+1)h)^{(\alpha-1)}_h f(sh),
\label{hSum}
\end{equation}
where $\alpha \ge 0$, $(_a\Delta^{0}_{h}f)(t)=f(t)$, $f$ is defined on
$(h\mathbb{N})_a$, and $_a\Delta^{-\alpha}_h$ on  
$(h\mathbb{N})_{a+\alpha h}$. $(h\mathbb{N})_t=\{t,t+h, t+2h, ...\}$.
The $h$-factorial $t^{(\alpha)}_h$ is defined as
\begin{equation}
t^{(\alpha)}_h =h^{\alpha}\frac{\Gamma(\frac{t}{h}+1)}{\Gamma(\frac{t}{h}+1-\alpha)}= h^{\alpha}\Bigl(\frac{t}{h}\Bigr)^{(\alpha)}, \ \ \frac{t}{h} \ne -1, -2, -3,
...
\label{hFrFac}
\end{equation}
With $m=\lceil \alpha \rceil$   the Riemann-Liouville (left) h-difference is defined as
{\setlength\arraycolsep{0.5pt}
\begin{eqnarray}
&&(_a\Delta^{\alpha}_h x)(t) =  (\Delta^{m}_h
(_a\Delta^{-(m-\alpha)}_{h}x))(t) \nonumber \\  
&&=\frac{h}{\Gamma(m-\alpha)} \Delta^m_h \sum^{\frac{t}{h}-(m-\alpha)}_{s=\frac{a}{h}}(t-(s+1)h)^{(m-\alpha-1)}_h 
x(sh)
\label{hFDRL}
\end{eqnarray}
}
and  the Caputo (left) h-difference is defined as
{\setlength\arraycolsep{0.5pt}
\begin{eqnarray}
&&(_a\Delta^{\alpha}_{h,*} x)(t) =  
(_a\Delta^{-(m-\alpha)}_h (\Delta^{m}_{h}x))(t) \nonumber \\  
&&=\frac{h}{\Gamma(m-\alpha)} \sum^{\frac{t}{h}-(m-\alpha)}_{s=\frac{a}{h}}(t-(s+1)h)^{(m-\alpha-1)}_h 
(\Delta^m_hx)(sh),
\label{hFDC}
\end{eqnarray}
}
where $(\Delta^{m}_{h}x))(t)$ is the $m$th power of the forward $h$-difference operator 
\begin{equation} 
(\Delta_{h}x)(t)=\frac{x(t+h)-x(t)}{h}.
\label{FHD}
\end{equation} 

As it has been noted in \cite{hdif1,hdif3,hdif3n}, due to the convergence of
solutions of fractional Riemann-Liouville h-difference equations 
when $h \rightarrow 0$ to 
solutions of the corresponding differential 
equations,  they can be used
to solve fractional Riemann-Liouville differential equations
numerically. A proof of the convergence (as $h \rightarrow 0$) of   
fractional Caputo h-difference operators to the corresponding  
fractional Caputo differential  operators for $0 < \alpha \le 1$ can be
found in \cite{hdif4n} (Proposition 17).

In what follows, we will consider fractional Caputo difference maps - the only fractional difference maps which behavior has been investigated.  
The following theorem \cite{DifSum,ME8,ME9,Fall} is essential to derive the universal fractional difference map:
\begin{theorem}
 For $\alpha \in \mathbb{R}$, $\alpha \ge 0$ the Caputo-like 
difference equation 
\begin{equation}
_0^C\Delta^{\alpha}_t x(t) = -G_K(x(t+\alpha-1)),
\label{LemmaDif_n}
\end{equation}
where $t\in \mathbb{N}_{m}$, with the initial conditions 
 \begin{equation}
\Delta^{k} x(0) = c_k, \ \ \ k=0, 1, ..., m-1, \ \ \ 
m=\lceil \alpha \rceil
\label{LemmaDifICn}
\end{equation}
is equivalent to the map with falling factorial-law memory
{\setlength\arraycolsep{0.5pt}   
\begin{eqnarray} 
&&x_{n+1} =   \sum^{m-1}_{k=0}\frac{\Delta^{k}x(0)}{k!}(n+1)^{(k)} 
\nonumber \\
&&-\frac{1}{\Gamma(\alpha)}  
\sum^{n+1-m}_{s=0}(n-s-m+\alpha)^{(\alpha-1)} 
G_K(x_{s+m-1}), 
\label{FalFacMap}
\end{eqnarray}
}
where $x_k=x(k)$, 
which is called the fractional difference Caputo universal  
$\alpha$-family of maps.
\label{T1}
\end{theorem}

To consider h-differences, we will extend this theorem using the property
(see \cite{hdif3}) 
\begin{equation}
(_0\Delta^{\alpha}_{h,*} x)(t)={h^{-\alpha}} _0^C\Delta^{\alpha}_t   \bar{x}\Bigl( \frac{t}{h} \Bigr),
\label{hNonh}
\end{equation}
where $x$ 
is defined on
$(h\mathbb{N})_a$,  $_a\Delta^{\alpha}_{h,*}$ on  
$(h\mathbb{N})_{a+\alpha h}$, and  $\bar{x}(s)=x(sh)$.
It is easy to show that the following theorem is a generalization of Theorem~\ref{T1}:

\begin{theorem}
 For $\alpha \in \mathbb{R}$, $\alpha \ge 0$ the Caputo-like 
h-difference equation 
\begin{equation}
(_0\Delta^{\alpha}_{h,*} x)(t) = -G_K(x(t+(\alpha-1)h)),
\label{LemmaDif_n_h}
\end{equation}
where $t\in (h\mathbb{N})_{m}$, with the initial conditions 
 \begin{equation}
(_0\Delta^{k}_h x)(0) = c_k, \ \ \ k=0, 1, ..., m-1, \ \ \ 
m=\lceil \alpha \rceil
\label{LemmaDifICn_h}
\end{equation}
is equivalent to the map with $h$-factorial-law memory
{\setlength\arraycolsep{0.5pt}   
\begin{eqnarray} 
&&x_{n+1} =   \sum^{m-1}_{k=0}\frac{c_k}{k!}((n+1)h)^{(k)}_h 
\nonumber \\
&&-\frac{h^{\alpha}}{\Gamma(\alpha)}  
\sum^{n+1-m}_{s=0}(n-s-m+\alpha)^{(\alpha-1)} 
G_K(x_{s+m-1}), 
\label{FalFacMap_h}
\end{eqnarray}
}
where $x_k=x(kh)$, 
which is called the $h$-difference Caputo universal  
$\alpha$-family of maps.
\label{T2}
\end{theorem}

In the case of integer $\alpha$ the fractional difference universal map converges to 
\begin{equation}
x_{n+1}=-G_K (x_n)  \  \ {\rm for} \  \ \alpha = 0, \  \ {\rm to} \  \ x_{n+1}=x_n-hG_K(x_n)  
\  \ {\rm for} \  \  \alpha = 1, 
\label{IntFrD01} 
\end{equation}
and for $\alpha=N = 2$, with $p_{n+1}=(x_{n+1}-x_n)/h $, to
\begin{equation}
\begin{array}{c}
\left\{
\begin{array}{lll}
p_{n+1}=p_n-h G_K(x_{n}), \ \ n \ge 1, \ \ p_1=p_0,
\\ 
x_{n+1} = x_{n}+h p_{n+1},  \ \ n \ge 0.
\end{array}
\right.
\end{array} 
\label{IntFrD2}
\end{equation} 
N-dimensional, with  $N \ge 2$, difference universal maps are investigated in \cite{ME8}. They are volume preserving (as well as the N-dimensional universal maps of Section~\ref{SecUFM}).

All the above considered universal maps in the case $\alpha=2$ yield the standard map if $G_K(x)=K\sin(x)$ (harmonic nonlinearity) and we'll call them 
the standard $\alpha$-families of maps.
When $G_K(x)=x-Kx(1-x)$ (quadratic nonlinearity) 
in the one-dimensional case all maps
yield the regular logistic map and we'll call them the logistic 
$\alpha$-families of maps.

\section{Periodic sinks and their stability}
\label{sec:3}

As in regular dynamics, the notion of universality and transition to chaos
in fractional dynamics is related to the dependence of the phase space 
structure of fixed and periodic points (sinks) on systems' parameters. 
Presence of power-law memory leads to some new features that appear in 
fractional dynamics.
\begin{itemize}
\item
{
In addition to the dependence on nonlinearity parameters, the phase space structure of fractional systems depends on a memory (an order of a fractional derivative) parameter.
}
\item
{Periodic points in fractional dynamics exist in the asymptotic sense. 
As it has been shown in \cite{ZSE}, effects of memory on the phase space 
structure of fractional systems of the order $\alpha \in (1,2)$  are 
similar to the effects of dissipation. But in fractional systems periodic 
sinks have their basins of attraction to which they themselves may not 
belong \cite{ME2,ME1,ME5}. In the latter case a trajectory that starts 
from a sink jumps out of the sink and may end up in a different sink.
}
\item
{
Evolution of systems with memory, in general, follows cascade of
bifurcations type trajectories (CBTT). Two examples of CBTT are presented 
in 
Fig.1.
As time (number of iterations $n$) increases,
the trajectory bifurcates and may end as a periodic sink
(Fig.~1~$a$) 
or as a chaotic trajectory 
(Fig.~1~$b$).   
}
\item
{
Not only the time of convergence of trajectories to the periodic sinks but also the way in which convergence occurs depends on the initial
conditions. As $n \rightarrow \infty$, all trajectories in 
Fig.~2
converge to the same period two ($T=2$) sink (as in 
Fig.~2~$c$), 
but for small 
values of initial conditions $x_0$ all
trajectories first converge to a $T=1$ trajectory which then 
bifurcates and turns into the $T=2$ sink converging to its limiting value. 
As $x_0$ increases, the bifurcation point $n_{bif}$ gradually evolves from 
the right to the left 
(Fig.~2~$a$).   
Ignoring this feature may result (as, e.g., in 
\cite{FallC,Fall}) 
in very messy bifurcation diagrams. 
}
\end{itemize}

\begin{figure}[!t]
\begin{center}
\includegraphics[width=0.9\textwidth]{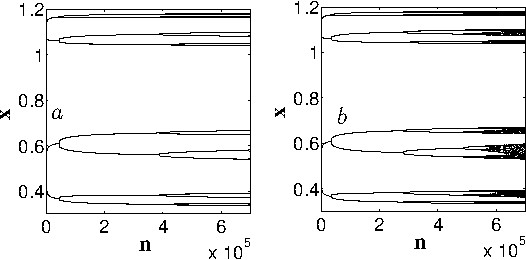}
\vspace{-0.25cm}
\caption{Two examples of cascade of bifurcations type trajectories in the
Caputo logistic $\alpha$-family of maps (Eq.(\ref{FrCMapx}) with $T=1$ and  
$G_K(x)=x-Kx(1-x)$ ) with $\alpha=0.1$ and $x_0=0.001$: 
(a) for the nonlinearity parameter $K=22.39$ the last bifurcation from period  
$T=16$ to period $T=32$ occurs after approximately $7 \times 10^5$ iterations; 
(b) when  $K=22.416$ the trajectory becomes chaotic after approximately 
$6 \times 10^5$ iterations.
}
\end{center}
\label{fig:1}
\end{figure}

\begin{figure}[!t]
\begin{center}
\includegraphics[width=0.9\textwidth]{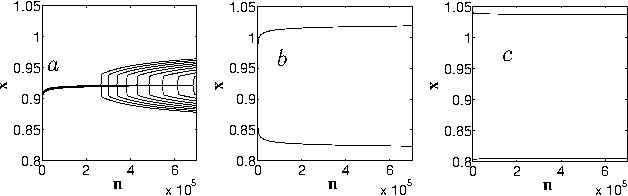}
\vspace{-0.25cm}
\caption{Asymptotically period two trajectories for the Caputo logistic
 $\alpha$-family of maps with  $\alpha=0.1$ and $K=15.5$: (a) nine
 trajectories with initial conditions $x_0$ from 0.29 (the rightmost
 bifurcation) to 0.37 (the leftmost bifurcation) with the step 0.01;
(b) $x_0=0.74$; (c) $x_0=0.94$.
}
\end{center}
\label{brT2}
\end{figure}

\subsection{$T=2$ sinks and stability of fixed points}
\label{T2section}

Stability of fixed points in fractional dynamical systems was investigated
in multiple publications, see, e.g., articles \cite{StA,StL,StM}, chapter
4 in book \cite{Petras}, and review \cite{StRev2013}; for stability in 
discrete fractional systems see, e.g., \cite{StB,ME1,StDis1}. There are
various ways to define stability and various methods and criteria 
to analyze it. In this paper we consider an asymptotic stability of
periodic points. A periodic point is asymptotically stable if there exists
an open set such that all trajectories with initial conditions from this set  
converge to this point as $t \rightarrow \infty$. It is known from the
study of the ordinary nonlinear dynamical systems that as a
nonliniearity parameter increases the system bifurcates. This means that
at the point (value of the parameter) of birth of the $T=2^{n+1}$ sink, the  
$T=2^{n}$ sink becomes unstable. In this section we will investigate the
$T=2$ sinks of discrete fractional systems and apply our results to
analyze stability of the systems' fixed points. As all published results on the existence and stability of the $T=2$ point were obtained for $h=1$, in this section we assume $h=1$.

All published results on the asymptotic stability of the stable fixed point and $T=2$ sink were obtained for the fractional and fractional difference standard and logistic $\alpha$-families of maps.

\subsubsection{Fractional standard map ($1<\alpha<2$)}
\label{alp1_2FSM}

First results on the first bifurcation and stability of the fixed point in discrete fractional systems were obtained in \cite{ME2,ME1,ME5} for the Riemann-Liouville standard $\alpha$-family of maps ($G_K(x)=K \sin(x)$)  for $1<\alpha<2$. In this case the map Eq.~(\ref{FrRLMapx}) can be written as a two-dimensional map considered on a cylinder 
\begin{equation} \label{FSMRLp}
p_{n+1} = p_n - K \sin x_n ,
\end{equation}
\begin{equation} \label{FSMRLx}
x_{n+1} = \frac{1}{\Gamma (\alpha )} 
\sum_{i=0}^{n} p_{i+1}V^1_{\alpha}(n-i+1) 
, \ \ \ \ ({\rm mod} \ 2\pi ) ,
\end{equation}
where 
\begin{equation} \label{V1}
V^k_{\alpha}(m)=m^{\alpha -k}-(m-1)^{\alpha -k} 
\end{equation}
and the momentum $p(t)$ is defined as
\begin{equation} \label{MomRL}
p(t)= \, _0D^{\alpha-1}_t x(t).
\end{equation} 
The Caputo standard $\alpha$-family of maps from Eq.~(\ref{FrCMapx}) can be considered on a torus and written as
\begin{equation} \label{FSMCp}
p_{n+1} = p_n 
-\frac{K}{\Gamma (\alpha -1 )} 
\Bigl[ \sum_{i=0}^{n-1} V^2_{\alpha}(n-i+1) \sin x_i 
+ \sin x_n \Bigr],\ \ ({\rm mod} \ 2\pi ), 
\end{equation}
\begin{equation} \label{FSMCx}
x_{n+1} = x_n + p_0 
-\frac{K}{\Gamma (\alpha)} 
\sum_{i=0}^{n} V^1_{\alpha}(n-i+1) \sin x_i,\ \ ({\rm mod} \ 2\pi ). 
\end{equation}
Both maps have the fixed point in the origin $(0,0)$. 
Numerical simulations show that both maps also have two $T=2$ sinks:
the antisymmetric sink, with
\begin{equation} \label{T2point} 
p_{n+1} = -p_n, \    \  x_{n+1} = -x_n,
\end{equation}
and the $\pi$-shift sink, with 
\begin{equation} \label{T2pointN} 
p_{n+1} = -p_{n}, \    \  x_{n+1} = x_n-\pi.
\end{equation}
For the Riemann-Liouville family of maps
there are two types of convergence
of the trajectories to the fixed point and the $T=2$ sinks: fast (from the basins of attraction) with
\begin{equation} \label{FastConv} 
\delta x_n \sim n^{-1-\alpha}, \  \ \delta p_n
\sim n^{-\alpha}
\end{equation}
and slow with  
\begin{equation} \label{SlowConv}
\delta x_n \sim n^{-\alpha}, \  \  \delta p_n
\sim n^{1-\alpha}.
\end{equation} 
For the Caputo family of maps
\begin{equation} \label{CaputoConv}
\delta x_n \sim n^{1-\alpha}, \  \  \delta p_n
\sim n^{1-\alpha}. 
\end{equation}

The antisymmetric $T=2$ sink $(x_l,p_l)$ and $(-x_l,-p_l)$,  Eq.~(\ref{T2point}), can be found considering the $n \rightarrow \infty$ limit in  Eqs.~(\ref{FSMRLp})~and~(\ref{FSMRLx}):
\begin{equation} \label{RLplim} 
p_l = \frac{K}{2} \sin(x_l),
\end{equation}
\begin{equation}\label{xl1} 
x_l = \frac{K}{2 \Gamma(\alpha)} V_{\alpha l} \sin(x_l),
\end{equation}
where
\begin{equation} \label{Val} 
 V_{\alpha l}  =  \sum_{k=1}^{\infty} (-1)^{k+1} V_{\alpha}^1(k).
\end{equation}
A high accuracy algorithm for calculating the slow converging series in Eq.~(\ref{Val}) can be found in the Appendix section of
\cite{ME4}.
Eq.~(\ref{xl1}) has a solution and the $T=2$ sink exists when    
\begin{equation} \label{Kcr1} 
K \ge K_{s1}(\alpha) = \frac{2 \Gamma(\alpha)}{V_{\alpha l}}.
\end{equation}
The opposite condition, as found in \cite{ME1}, is the condition of the stability of the $(0,0)$ fixed point. The same condition can be shown for the Caputo standard $\alpha$-family of maps. It is used to plot the part $1<\alpha \le2$ of the bottom thin line in 
Fig.~5~$a$,
which is a two-dimensional ($\alpha-K$) bifurcation diagram. The fixed point (0,0) is stable below this line.   

$\pi$-shift $T=2$ sink 
\begin{equation} \label{T2nonAS} 
p_{n} = (-1)^np_l, \    \  x_{n} = x_l-\frac{\pi}{2}[1-(-1)^n]
\end{equation}
can be found plugging the asymptotic expression for $x_n$ from Eq.~(\ref{T2nonAS}) and
\begin{equation} \label{T2nonLimAS} 
p_{n} = (-1)^np_l+An^{1-\alpha}
\end{equation}
into Eqs.~(\ref{FSMRLp})~and~(\ref{FSMRLx}) and considering 
$n \rightarrow \infty$ limit.
This gives
\begin{equation} \label{T2nonASA} 
 p_l = K/2 \sin(x_l), \    \ A= \frac{2 x_l-\pi}{2 \Gamma(2-\alpha)}, 
\end{equation}
\begin{equation} \label{T2nonASxsol} 
\sin(x_l)= \frac{\pi \Gamma(\alpha)}{K V_{\alpha l}}, 
\end{equation}
which has solutions for 
\begin{equation} \label{T2nonASKc} 
K>K_{s2}= \frac{\pi \Gamma(\alpha)}{V_{\alpha l}}=\frac{\pi}{2}K_{s1}. 
\end{equation}
$K_{s2}(\alpha)$ is used to plot the part $1<\alpha \le2$ of the middle thin line in 
Fig.~5~$a$.

\subsubsection{Fractional logistic map ($1 < \alpha \le2$)}
\label{alp1_2LSM}

A fractional generalization of the logistic map became possible after a small time delay was introduced into the differential equation describing a periodically kicked system Eq.~(\ref{UM1D2Ddif}) (see \cite{ME4}). The logistic Riemann-Liouville $\alpha$-family of maps ($G_K(x)=x-Kx(1-x)$) can be written as
{\setlength\arraycolsep{0.5pt}
\begin{eqnarray}
\label{FLMRLp}
&&p_{n+1} = p_n + K x_n (1-x_n)-x_n ,  \\
&&x_{n+1} = \frac{1}{\Gamma (\alpha )} 
\sum_{i=0}^{n} p_{i+1}V^1_{\alpha}(n-i+1).
\label{FLMRLx}
\end{eqnarray}
}
Numerical simulations show that for $0<K \le 1$ all converging trajectories converge to the fixed point $(0,0)$ as 
$x \sim n^{-\alpha -1}$, $p \sim n^{-\alpha}$. For $1<K<K_{l1}$ 
the only stable periodic sink is the fixed point $((K-1)/K,0)$. The rate of convergence to this fixed point is  $\delta x \sim n^{-\alpha}$, $p \sim n^{-\alpha+1}$.
At $K=K_{l1}$ the fixed point becomes unstable and the stable antisymmetric in $p$  period two sink appears. 
From the results of numerical simulations \cite{ME4}, the asymptotic behavior of converging to the $T=2$ sink trajectories follows the power law
\begin{equation} \label{P2Asy} 
p_n=p_l(-1)^n+ \frac{A}{n^{\alpha -1}} .\ \  
\end{equation}
Substituting this expression for $p_n$ into Eq.~(\ref{FLMRLx}) and considering even values of $n$, we obtain
{\setlength\arraycolsep{0.5pt}
\begin{eqnarray}
&&x_{lo}=\lim_{n \rightarrow \infty}x_{2n+1}=\frac{p_l}{\Gamma(\alpha)} 
\lim_{n \rightarrow \infty}\sum^{2n+1}_{k=1}(-1)^k V_{\alpha}^1(k) \nonumber \\ 
&&\hspace{-0.2cm}+\frac{A}{\Gamma(\alpha)}\lim_{n \rightarrow \infty}
\sum^{2n-1}_{k=1}\frac{\alpha-1}{k^{\alpha-1}(2n-k)^{2-\alpha}}=
-\frac{p_l}{\Gamma(\alpha)}V_{\alpha l}  
\label{limxo} \\
&&\hspace{-0.2cm}+\frac{(\alpha-1)A}{\Gamma(\alpha)} \int^{1}_0
\frac{x^{1-\alpha} dx}{(1-x)^{2-\alpha}}
=-\frac{p_l}{\Gamma(\alpha)}V_{\alpha l}+A \Gamma(2-\alpha). \nonumber
\end{eqnarray}
}
Here we took into account that the sum on the second line of the last equation is the Riemann sum for the integral on the third line, which is equal to the Beta-function 
$B(2-\alpha,\alpha-1)=\Gamma(2-\alpha)\Gamma(\alpha-1)$.
 Similarly, 
\begin{equation} \label{limxe} 
x_{le}=\lim_{n \rightarrow \infty}x_{2n}=\frac{p_l}{\Gamma(\alpha)}V_{\alpha l}+A \Gamma(2-\alpha).
\end{equation}
Then, in the limit $n \rightarrow \infty$,   Eq.~(\ref{FLMRLp}) gives
{\setlength\arraycolsep{0.5pt}
\begin{eqnarray}
\label{limpo}
&&-2p_{l}=Kx_{le}(1-x_{le})-x_{le} ,  \\
&&2p_{l}=Kx_{lo}(1-x_{lo})-x_{lo}.
\label{limpe}
\end{eqnarray}
}
Two fixed points, $x_{lo}=x_{le}=p_{l}=A=0$ 
and $x_{lo}=x_{le}=x_l=(K-1)/K$, $p_{l}=0$, $A=x_l/\Gamma (2-\alpha)$,
are the two expected solutions of the system of four equations Eqs.~(\ref{limxo})-(\ref{limpe}). For the two remaining solutions with 
$x_{lo} \ne x_{le}$
\begin{equation} \label{A} 
A= \frac{K-1+\frac{2\Gamma(\alpha)}{V_{\alpha l}}}{2K\Gamma(2-\alpha)}
\end{equation}
and the quadratic equation defining $x_{le}$ and $x_{lo}$ can be written as 

\begin{equation}
x_{lo,le}^2-\Bigl( \frac{2\Gamma(\alpha)}{V_{\alpha l}K} 
+\frac{K-1}{K}  \Bigr)x_{lo,le} +
\frac{2\Gamma^2(\alpha)}{(V_{\alpha l}K)^2}
+\frac{(K-1)\Gamma(\alpha)}{V_{\alpha l}K^2}=0.
\label{eqT2logn}
\end{equation}
The solutions of this equation  
\begin{equation}
x_{lo,le}=\frac{K_{s1}+K-1 \pm \sqrt{(K-1)^2-K_{s1}^2}}{2K}
\label{eqT2logSolun}
\end{equation}
are defined when
\begin{equation}
K \ge 1 + \frac{2\Gamma(\alpha)}{V_{\alpha l}}=1+K_{s1} \ \ {\rm or} \ \
K \le 1 - \frac{2\Gamma(\alpha)}{V_{\alpha l}}=1-K_{s1}.
\label{condT2logN}
\end{equation}
From  $V_{\alpha l}<1$ and $\Gamma(\alpha)>0.885$ for $\alpha>0$
follows that $ 2\Gamma(\alpha)/V_{\alpha l}>1$ and, considering 
only $K>0$,  we may ignore the
second of the inequalities in Eq.~(\ref{condT2logN}). 
We may also note that
the fixed point $x=(K-1)/K$ is stable when
\begin{equation}
1 \le K <K_{l1}= 1 + \frac{2\Gamma(\alpha)}{V_{\alpha l}}=1+K_{s1}.
\label{condT2lognN}
\end{equation} 
$K_{l1}(\alpha)$ is used to plot the part $1<\alpha \le2$ of the bottom thin line in 
Fig.~5~$b$.

\subsubsection{Fractional and fractional difference standard $\alpha$-families of maps for $0<\alpha<1$}

In this and the next sections we will follow the results obtained 
in \cite{ME8}.
For $0<\alpha<1$ fractional and fractional difference maps, Eq.~(\ref{FrCMapx}) and Eq.~(\ref{FalFacMap_h}), can be written (with $h=1$ and $G_K(x)=K \sin(x)$) in the universal form
\begin{equation}
x_{n}=  x_0- 
\frac{K}{\Gamma(\alpha)}\sum^{n-1}_{k=0} W_{\alpha}(n-k) \sin{(x_k)},
\label{1_2DSinMaps}
\end{equation}
where $W_{\alpha}(s) = s^{\alpha-1}$ for the fractional map and
 $W_{\alpha}(s) ={\Gamma(s+\alpha-1)}/{\Gamma(s)}$ for the fractional
 difference map.

For $\alpha=0$ the Caputo fractional standard map is identically zero and fractional difference is the sine map 
\begin{equation}
x_{n+1} = -K\sin(x_n), \ \ \ ({\rm mod} \ 2\pi ).
\label{SM0D}
\end{equation} 
For $\alpha=1$ both maps converge to the circle map with zero driven phase
\begin{equation}
x_{n+1}= x_n + K \sin (x_n), \ \ \ \ ({\rm mod} \ 2\pi ).
\label{SM1DNeg} 
\end{equation}
In the sine map and in the circle map with zero driven phase,
when the $x=0$ sink becomes unstable, it bifurcates into
the symmetric $T=2$ sink in which $x_{n+1}=-x_{n}$.
Following the results of \cite{ME8}, let's assume that 
this property persists (asymptotically) for $\alpha \in (0,1)$.
Eq.~(\ref{1_2DSinMaps}) can be written as
{\setlength\arraycolsep{0.5pt}   
\begin{eqnarray} 
&&x_{n+1}=  x_n- 
\frac{K}{\Gamma(\alpha)} \Bigl\{  W_{\alpha}(1) \sin{(x_n)} \nonumber \\
&&+\sum^{n-1}_{k=0} \sin{(x_k)} [W_{\alpha}(n-k+1)-W_{\alpha}(n-k)]\Bigr\}. 
\label{1_2DSinMapsD}
\end{eqnarray}
}
Because $W_{\alpha}(n-k+1)-W_{\alpha}(n-k) \rightarrow 0$
when $n \rightarrow \infty$, substituting $j=n-k$, asymptotically for large $n$ Eq.~(\ref{1_2DSinMapsD}) can be written as
{\setlength\arraycolsep{0.5pt}   
\begin{eqnarray} 
&&x_n=\frac{K}{2\Gamma(\alpha)}\Bigl\{
  W_{\alpha}(1) +\sum^{\infty}_{j=1} (-1)^j[W_{\alpha}(j+1)-W_{\alpha}(j)]\Bigr\}\sin{(x_n)},
\label{1_2DSinMapsD1} 
\end{eqnarray}
}
where the alternating series on the right side converges because
its terms converge to 0 monotonically.
This equation has real non-trivial solutions when 
{\setlength\arraycolsep{0.5pt} 
\begin{eqnarray} 
&&K>K_{s1} =\frac{2\Gamma(\alpha)}{W_{\alpha}(1)
+\sum^{\infty}_{j=1} (-1)^j[W_{\alpha}(j+1)-W_{\alpha}(j)]}.
\label{1_2DSinMapsKcr}
\end{eqnarray}
}
This expression for $K_{s1}$ is used to plot the parts $0<\alpha \le1$ of the bottom thin and bold lines in 
Fig.~5~$a$.

When the symmetric $T=2$ sink becomes unstable it gives birth to
the $\pi$-shift $T=2$ sink in which $|x_{n+1}-x_{n}|=\pi$. The asymptotic analysis similar to the above performed for the symmetric $T=2$ sink yields the following equation to define the asymptotic values for this sink
{\setlength\arraycolsep{0.5pt}   
\begin{eqnarray} 
&&\pm \pi=\frac{K}{\Gamma(\alpha)}\Bigl\{
  W_{\alpha}(1) +\sum^{\infty}_{j=1} (-1)^j[W_{\alpha}(j+1)-W_{\alpha}(j)]\Bigr\}\sin{(x_n)},
\label{1_2DSinMapsD2} 
\end{eqnarray}
}
which has solutions when
\begin{equation} 
K>K_{s2}=\pi K_{s1}/2.
\label{Kcr2}
\end{equation}
This expression for $K_{s2}$ is used to plot the part $0<\alpha \le1$ of the middle thin line in 
Fig.~5~$a$.
It also could be used to calculate the middle bold line, but in this paper we use the results of the direct numerical calculations \cite{ME8} instead. The difference between the results of the direct numerical simulations and the calculations using Eq.~(\ref{Kcr2}) is evident when $\alpha < 0.15$.
This difference is due to the slow, as $n^{-\alpha}$,
convergence of trajectories.

\subsubsection{Fractional difference standard $\alpha$-families of maps for $1<\alpha<2$}

When $1<\alpha<2$ the fractional difference Caputo standard 
$\alpha$-FM Eq.~(\ref{FalFacMap_h}) can be written as a 
two-dimensional map \cite{ME8}
{\setlength\arraycolsep{0.5pt}   
\begin{eqnarray} 
&&p_{n} =  p_1 -\frac{K}{\Gamma(\alpha-1)}
\times \sum^{n}_{s=2}\frac{\Gamma(n-s+\alpha-1)}
{\Gamma(n-s+1)}\sin (x_{s-1}) 
, \ \ ({\rm mod} \ 2\pi ),
\label{SMgt1p} \\
&& x_n=x_{n-1}+p_n, \ \ ({\rm mod} \ 2\pi ),  \ \ n \ge 1,
\label{SMgt1x}
\end{eqnarray}
}
where $p_n=\Delta x_{n-1}=x_n-x_{n-1}$.
 
As in the fractional standard map, in the fractional difference standard map, when the $(0,0)$ fixed point becomes unstable,
it bifurcates into the $T=2$ antisymmetric sink $x_{n+1}=-x_n$,
$p_{n+1}=-p_n$, which later, at $K$ for which
$x_n=\pi/2$, turns into two $\pi$-shift $T=2$ sinks \cite{ME8}.
For the anti-symmetric sink, in the limit $n \rightarrow \infty$ Eq.~(\ref{SMgt1x}) yields $p_n=2x_n$ and Eq.~(\ref{SMgt1p})
 {\setlength\arraycolsep{0.5pt}   
\begin{eqnarray} 
&&p_n=\frac{K}{2\Gamma(\alpha-1)}\Bigl\{
  W_{\alpha-1}(1)+\sum^{\infty}_{j=1} (-1)^j[W_{\alpha-1}(j+1)-W_{\alpha-1}(j)]\Bigr\}\sin{(x_n)},
\label{1_2DSinMapsD1P} 
\end{eqnarray}
}
where, as in Eq~(\ref{1_2DSinMaps}),
$W_{\alpha}(s) ={\Gamma(s+\alpha-1)}/{\Gamma(s)}$.
The equations for
the antisymmetric $T=2$ sink $(x_n,p_n)$ are
{\setlength\arraycolsep{0.5pt}   
\begin{eqnarray} 
&&x_n=\frac{K}{4\Gamma(\alpha-1)}\Bigl\{
  W_{\alpha-1}(1) +\sum^{\infty}_{j=1} (-1)^j[W_{\alpha-1}(j+1)-W_{\alpha-1}(j)]\Bigr\}\sin{(x_n)},
\label{1_2DSinMapsD1Xnnn} \\
&&p_n=2x_n.
\label{1_2DSinMapsD1XPnn}
\end{eqnarray}
}
They have solutions for
\begin{equation}
K>K_{s1}(\alpha)=2K_{s1}(\alpha-1),
\label{KCR1_2}
\end{equation}
where $K_{c1d}(\alpha-1)$ is defined by  Eq.~(\ref{1_2DSinMapsKcr}). This
result is used to plot the part $1<\alpha \le2$ of the bottom bold line 
in Fig.~5~$a$. 

Equations defining the $\pi$-shift $T=2$ sink can be written as 
{\setlength\arraycolsep{0.5pt}   
\begin{eqnarray} 
&&\pm \pi=\frac{K}{2\Gamma(\alpha-1)}\Bigl\{
  W_{\alpha-1}(1) +\sum^{\infty}_{j=1} (-1)^j[W_{\alpha-1}(j+1)-W_{\alpha-1}(j)]\Bigr\}\sin{(x_n)},
\label{1_2DSinMapsD1Xn} \\
&&p_n=\pm \pi.
\label{1_2DSinMapsShift}
\end{eqnarray}
}
$\pi$-shift sink exists when 
\begin{equation}
K>K_{s2}(\alpha)=\frac{\pi}{2}K_{s1}(\alpha).
\label{KCR1_2Shift}
\end{equation}
This
result is used to plot the part $1<\alpha \le2$ of the middle bold line 
in Fig.~5~$a$.

The fixed point in the origin is stable for $K<K_{s1}$ and the convergence of trajectories to the fixed point follows the power law 
$x_n \sim n^{1-\alpha}$ and $p_n \sim n^{-\alpha}$.

\subsection{$T=2^n$ sinks}
\label{T2TonSection}

Investigation of the $T=2^n$-sinks' stability with $n>2$ by analytic methods is complicated.  In papers \cite{ME2,ME3,ME4,ME6,ME7,ME8,ME9,Chaos2015,ME1,ME5} this is done by numerical simulations on individual trajectories with various values of parameters ($K$ and $\alpha$) and initial conditions. As in the case of the fixed point and $T=2$-sink, stability of the high order sinks is asymptotic. Trajectories, which converge to $T=2^n$-sinks ($n^{th}$ order sinks) stable in the limit $n \rightarrow \infty$, may first converge to low order sinks and then, through cascades of period doubling bifurcations, converge to the sinks of the $n^{th}$ order. Cascade of bifurcations type trajectories are the fundamental features of the discrete fractional systems. Their presence makes drawing of various kinds of fractional bifurcation diagrams (the subject of the next section) difficult. Fractional bifurcation diagrams strongly depend on the number of iterations and initial conditions of individual trajectories used in the analysis. The larger the number of iterations used in the calculations, the closer the calculated values of the sinks to their limiting values.

\section{Fractional bifurcation diagrams}
\label{sec:4}

Fractional maps demonstrate the universal scenario of transition to chaos through the period doubling cascade of bifurcations with the change in a nonlinearity parameter, similar to the one described in \cite{May}. This is illustrated in 
Fig.~3, 
in which the bifurcation diagrams ($x$ vs. $K$) for various considered families of maps with $\alpha=0.8$ are presented. Compared to the integer case $\alpha=1$, bifurcation diagrams for the fractional maps are stretched along the $K$ axis while bifurcation diagrams for the fractional difference maps are contracted.
The existence of self-similarity,
corresponding constants (analogs of the Feigenbaum constants), and
their dependence on $\alpha$ in fractional maps are not investigated.
\begin{figure}[!t]
\begin{center}
\includegraphics[width=0.9\textwidth]{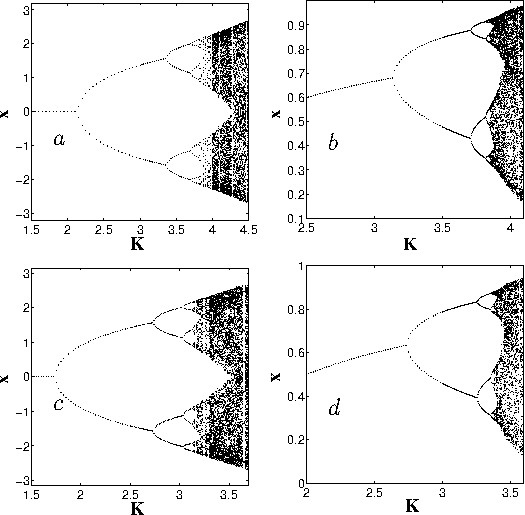}
\vspace{-0.25cm}
\caption{The bifurcation diagrams for fractional Caputo standard (a) 
and logistic (b) maps and for fractional difference Caputo standard
(c) and logistic (d) maps with $\alpha=0.8$.   
}
\end{center}
\label{BD}
\end{figure}
The dependence of the bifurcation diagrams on the number of iterations is demonstrated in Fig.~\ref{BifLT1shift}. Bifurcation diagrams obtained after five thousand iteration look much nicer than those obtained after two hundred. But, as it follows from 
Figs.~1~and~2, 
even 5000 iterations are not enough for computation of the asymptotic bifurcation diagrams. 

The two-dimensional bifurcation diagrams 
Fig.~5 
are obtained by combining the results of the computations of the bifurcation ($x$ vs. $K$) diagrams after 5000 iterations for fixed values of $\alpha$. Because for small $\alpha$ convergence of trajectories to their asymptotic values is very slow, the results in the diagrams for $\alpha<0.15$ do not represent well the asymptotic values and can be improved in future.    
\begin{figure}[!t]
\centering
\includegraphics[width=0.5\textwidth]{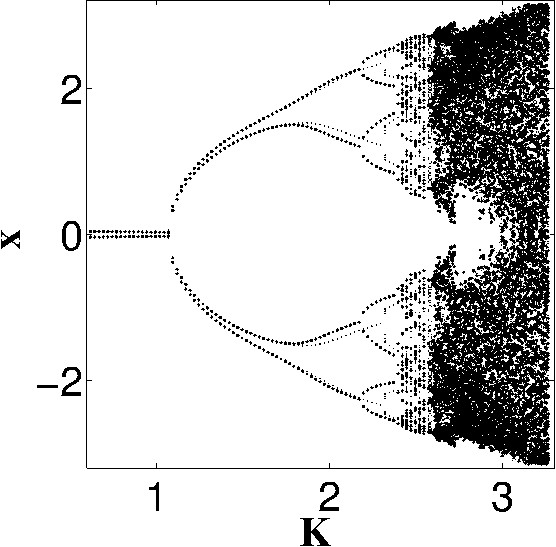}
\caption{Two bifurcation diagrams for the  fractional difference Caputo Standard
$\alpha$FM with   $\alpha=0.1$ and  $x_0=0.1$ calculated after 200
iterations (regular points) and 5000 iterations (bold points). This figure is reprinted from \cite{ME8}, with the permission of AIP Publishing.
}
\label{BifLT1shift}
\end{figure}
\begin{figure}[!t]
\begin{center}
\includegraphics[width=0.9\textwidth]{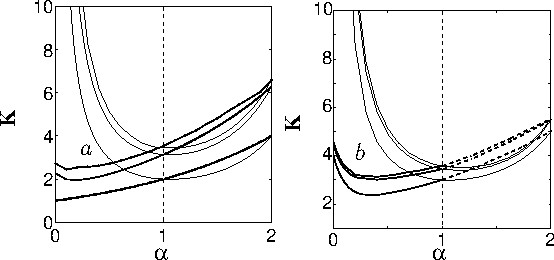}
\vspace{-0.25cm}
\caption{
2D bifurcation diagrams for fractional (thing lines) and
fractional difference (bold lines) Caputo standard (a) 
and logistic (b) maps. First bifurcation, transition from the stable fixed 
point to the stable period two ($T=2$) sink, 
occurs on the bottom curves.
$T=2$ sink (in the case of the standard $\alpha$-families of maps,
antisymmetric T=2 sink with $x_{n+1}=-x_n$)
is stable between the bottom and the middle curves. Transition 
to chaos occurs on the top curves. Period doubling bifurcations leading to chaos occur in the narrow band between the middle and the top curves.
The bottom curves in (a) are obtained using 
Eqs.~(\ref{Kcr1}),~(\ref{1_2DSinMapsKcr}),~and~(\ref{KCR1_2}).
The thin middle curve in (a) is obtained using 
Eqs.~(\ref{T2nonASKc})~and~(\ref{Kcr2}).
The $1 < \alpha \le 2$ part of the middle bold line in (a) is obtained using Eq.~(\ref{KCR1_2Shift}).
The $1 < \alpha \le 2$ part of the bottom thin line in (b) is obtained using Eq.~(\ref{condT2lognN}).
The remaining curves, except the bold dashed curves in (b), are results of the direct numerical simulations. The bold dashed curves in (b) are obtained by interpolation. 
}
\end{center}
\label{BD2d}
\end{figure} 

Looking at the 2D 
bifurcation diagrams one may note that systems with power-law memory should demonstrate 
bifurcations with changes in the memory parameter $\alpha$ when the
nonlinearity parameter $K$ stays constant. This property of
systems with power-law memory is demonstrated in 
Fig.~6 and
it may explain how changes/failures in live biological species can be
caused by changes in their memory and nervous system. This also may  
explain how some diseases may be treated by treating
the nervous system.  
\begin{figure}[!t]
\begin{center}
\includegraphics[width=0.75\textwidth]{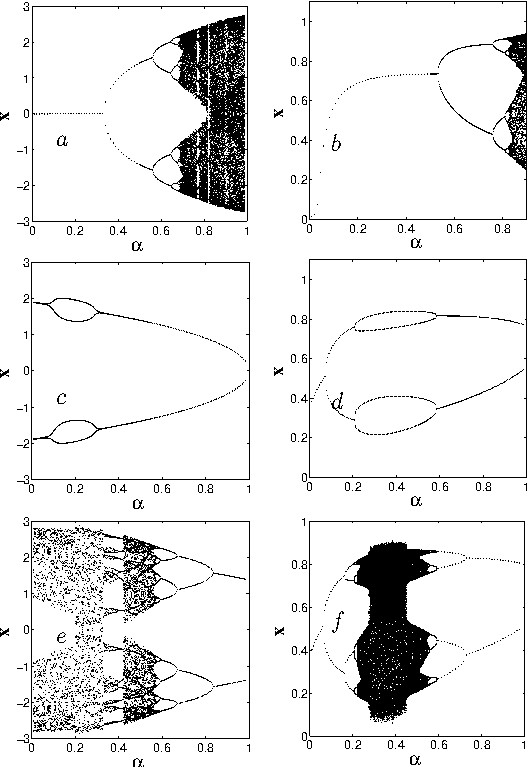}
\vspace{-0.25cm}
\caption{The memory $\alpha$-bifurcation diagrams for fractional Caputo 
standard (a) 
and logistic (b) maps and for fractional difference Caputo standard
(c) and (e) and logistic (d) and (f) maps obtained after 5000 iteration.
$K=4.2$ in (a), $K=3.8$ in (b), $K=2.0$ in (c), $K=3.1$ in (d), $K=2.8$ in (e),
and   $K=3.2$ in (7).   
}
\end{center}
\label{BDalp}
\end{figure}

\section{Conclusion}
\label{sec:5}

The following citation from Wikipedia, ``universality is the observation that there are properties for a large class of systems that are independent of the dynamical details of the system'', defines the notion of the universality in dynamical systems.
The universality in systems with power-law memory goes beyond the period
doubling with changes in nonlinearity and memory parameters and the
universal scenario of transition to chaos. Individual trajectories 
of such systems also demonstrate cascade of bifurcations type behavior.
In regular dynamics the universality has a mathematical expression in the form of the Feigenbaum function and constants.
This is only the beginning of the research on fractional universality and most of the results are obtained by numerical simulations. Those results introduce more questions than answers. Some of those questions are:  
\begin{itemize}
\item
{
What is the nature and the corresponding analytic description of the
bifurcations on a single trajectory of a fractional system?   
}
\item
{
What kind of self-similarity can be found in CBTT?   
}
\item
{
How to describe a self-similar behavior corresponding to the bifurcation diagrams of fractional systems? Can constants, similar to the Feigenbaum constants be found?   
}
\item
{
Can cascade of bifurcations type trajectories be found in continuous systems?   
}
\end{itemize}
Behavior of fractional systems at low values of $\alpha$ ($0<\alpha<0.15$) is very important in biological applications but is not well established and requires an additional investigation. 
\begin{figure}[!t]
\centering
\includegraphics[width=4in]{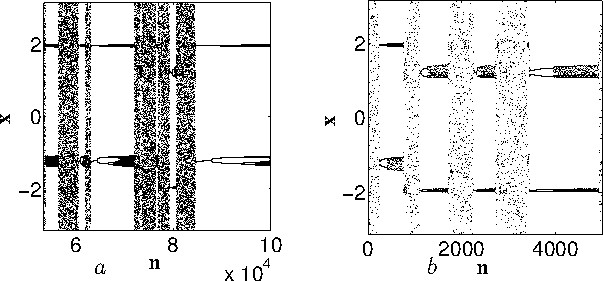}
\caption{Intermittent cascade of bifurcations type trajectories 
for the Caputo ($a$) and fractional difference Caputo 
 ($b$) 
Standard $\alpha$FMs. In ($a$) $\alpha=1.65$, $K=4.5$, $x_0=0.$, and
$p_0=0.3$. In ($b$) $\alpha=1.5$, $K=4.82$, $x_0=0.$, and
$p_0=0.01$.
This figure is reprinted from \cite{ME9}, with the permission of 
L\&H Scientific Publishing.
}
\label{F4}
\end{figure}

As mentioned in the introduction, there is a possibility for multiple applications of the fractional universality in biology. The human body is a system with power-law memory, which implies the possibility of medical applications. 
Fig.~2~$a$ 
suggests that, assuming some distribution (e.g., uniform) of the initial conditions of an asymptotically $T=2$ system with power-law memory, it is possible to calculate the probability distribution of times before the stable fixed point behavior of the system bifurcates. Comparison of probability distributions for various values of $K$ and $\alpha$ to the statistics of 
the times before sudden changes (e.g., deaths) after serious surgeries (e.g., heart transplants) may help to understand the state of a human body after the surgery and suggest some remedies.

The intermittent cascade of bifurcations type behavior is typical for systems with power-law memory (see Fig.~\ref{F4}). May this intermittency explain the intermittent behavior, transitions from stability to chaos and back to stability, in various socio-economic systems, which are systems with power-law memory? May the history of human society, with
repeating periods of dictatorship, democracy, and chaos, be modeled by the equations with power-law memory? There are many questions related to the topic of this review which motivate research on systems with power-law memory.


\begin{acknowledgement}  
The author expresses his gratitude to R. Cole and R. V. Kohn, 
for the opportunity to complete this work at the Courant Institute. 
The author is grateful to the organizers of the 6th International Conference on Nonlinear Science and Complexity in Sao Jose dos Campos, Brazil, for financial support. The author acknowledges continuing support from Yeshiva University.
\end{acknowledgement}

\end{document}